# An Integrated Approach for Identifying Relevant Factors Influencing Software Development Productivity

Adam Trendowicz[1], Michael Ochs[1], Axel Wickenkamp[1], Jürgen Münch[1], Yasushi Ishigai[2, 3], Takashi Kawaguchi[4]

[1] Fraunhofer IESE, Fraunhofer-Platz 1, 67663 Kaiserslautern, Germany
{trend, ochs, wicken, muench}@iese.fraunhofer.de

[2] IPA-SEC, 2-28-8 Honkomagome, Bunkyo-Ku, Tokyo, 113-6591, Japan
ishigai@ipa.go.jp

[3] Research Center for Information Technology Mitsubishi Research Institute, Inc.
3-6, Otemachi 2-Chome, Chiyoda-Ku, Tokyo, 100-8141, Japan
ishigai@mri.co.jp

[4] Toshiba Information Systems (Japan) Corporation, 7-1 Nissin-Cho,
Kawasaki-City 210-8540, Japan
kawa@tjsys.co.jp

**Abstract.** Managing software development productivity and effort are key issues in software organizations. Identifying the most relevant factors influencing project performance is essential for implementing business strategies by selecting and adjusting proper improvement activities. There is, however, a large number of potential influencing factors. This paper proposes a novel approach for identifying the most relevant factors influencing software development productivity. The method elicits relevant factors by integrating data analysis and expert judgment approaches by means of a multi-criteria decision support technique. Empirical evaluation of the method in an industrial context has indicated that it delivers a different set of factors compared to individual data- and expert-based factor selection methods. Moreover, application of the integrated method significantly improves the performance of effort estimation in terms of accuracy and precision. Finally, the study did not replicate the observation of similar investigations regarding improved estimation performance on the factor sets reduced by a data-based selection method.

## 1  Introduction

Many software organizations are still proposing unrealistic software costs, work within tight schedules, and finish their projects behind schedule and budget, or do not complete them at all [23]. This illustrates that reliable methods to manage software development effort and productivity are a key issue in software organizations. One essential aspect when managing development effort and productivity is the large number of associated and unknown influencing factors (so-called *productivity factors*) [27]. Identifying the right productivity factors increases the effectiveness of productivity improvement strategies by concentrating management activities directly on those development processes that have the greatest impact on productivity. On the other hand, focusing measurement activities on a limited number of the most relevant factors (goal-oriented measurement) reduces the cost of quantitative project management (collecting, analyzing, and maintaining the data). The computational complexity of numerous quantitative methods grows exponentially with the number of input factors [4], which significantly restricts their industrial acceptance.

In practice, two strategies to identify relevant productivity factors, promoted in the related literature, are widely applied. In *expert-based* approaches, one or more software experts decide about a factor's relevancy [24]. In *data-based* approaches, existing measurement data, covering a certain initial set of factors, are analyzed to identify a subset of factors relevant with respect to a certain criterion [6, 10]. These factor selection strategies have, however, significant practical limitations when applied individually. Experts usually base their decisions on subjective preferences and experiences. In consequence, they tend to disagree largely and omit relevant factors while selecting the irrelevant ones [24]. The effectiveness of data-based methods largely depends on the quantity and quality of available data. They cannot identify a relevant factor if it is not present in the initial (input) set of factors. Moreover, data analysis techniques are usually sensitive to messy (incomplete and inconsistent) data. Yet, assuring that all relevant factors are covered by a sufficient quantity of high-quality measurement data is simply not feasible in practice.

In this paper, we propose an integrated approach to selecting relevant productivity factors for the purpose of software effort estimation. We combine expert- with data-based factor selection methods, using a novel multi-criteria decision aid method called *AvalOn*. The presented approach is then evaluated in the context of a large software organization.

The remainder of the paper is organized as follows. Section 2 provides an overview of factor selection methods. Next, in Section 3, we present the integrated factor selection method, followed by the design of its empirical evaluation (Section 4) and an analysis of the results (Section 5). The paper ends with conclusions and further work perspectives (Section 6).

## 2 Related Work

*Factor selection* can be defined as the process of choosing a subset of M factors from the original space of N factors (M≤N), so that the factor space is optimally reduced according to a certain *criterion*. In principle, the selection process may be based on data analysis, expert assessments, or on both, experts and data. In *expert-based factor selection*, the factor space N is practically infinite and not known *a priori*. One or more experts may simply select a subset of relevant factors and/or weight factors with respect to their relevancy (e.g., using the *Likert scale* [21]). Factor ranking is equal to assigning discrete weights to them. In *data-based factor selection*, the factor space N is known and determined by available measurement data. Most of the factor selection methods originate from the data mining domain and belong to the so-called *dimensionality reduction* methods [8]. In principle, data-based selection methods assign weights to available factors. Weight may be dichotomous (factor selection), discrete (factor raking), or continuous (factor weighting).

The purpose of factor selection methods in software effort estimation is to reduce a large number of potential productivity factors (cost drivers) in order to improve estimation performance while maintaining (or reducing) estimation costs. Moreover, information on the most relevant influence factors may be used to guide measurement and improvement initiatives. In practice (authors' observation), relevant cost drivers are usually selected by experts and the selection process is often limited to uncritically adopting factors published in the related literature. Software practitioners adopt the complete effort model along with the integrated factors set (e.g., COCOMO [3]) or build their own model on factors adapted from an existing model. In both situations, they risk collecting a significant amount of potentially irrelevant data and getting limited performance of the resulting model.

In the last two decades, several data-based approaches have been proposed to support software organizations that are already collecting data on arbitrarily selected factors in selecting relevant factors. Various data analysis techniques were proposed to simply reduce the factor space by excluding potentially irrelevant ones (factor selection). The original version of the ANGEL tool [19] addressed the problem of optimal factor selection by exhaustive search. However, for larger factor spaces (>15-20), analysis becomes computationally intractable due to its exponential complexity. Alternative, less computationally expensive factor selection methods proposed in the literature include Principal Component Analysis (PCA) [22], Monte Carlo simulation (MC) [12], general linear models (GLM) [13], and wrapper factor selection [10, 6]. The latter approach was investigated using various evaluation models (e.g., regression [6], case-based reasoning [10]), and different search strategies (forward selection [6, 10], as well as random selection and sequential hill climbing [10]). In all studies, significant reduction (by 50%-75%) of an initial factors set and improved estimation accuracy (by 15%-379%) were reported. Chen et al. [6] conclude, however, that despite substantial improvements in estimation accuracy, removing more than half of the factors might not be wise in practice, because it is not the only decision criterion.

An alternative strategy to removing irrelevant factors would be assigning weights according to a factor's relevancy (*factor weighting*). The advantage of such an approach is that factors are not automatically discarded and software practitioners obtain information on the relative importance of each factor, which they may use to decide about the selection/exclusion of certain factors. Auer et al. [1] propose an optimal weighting method in the context of the k-Nearest Neighbor (k-NN) effort estimator; however, exponential computational complexity limits its practical applicability for large factor spaces. Weighting in higher-dimensionality environments can be, for instance, performed using one of the heuristics based on rough set analysis, proposed recently in [11]. Yet, their application requires additional overhead to discretize continuous variables.

A first trial towards an integrated factor selection approach was presented in [2], where Bayesian analysis was used to combine the weights of COCOMO II factors based on human judgment and regression analysis. Yet, both methods were applied on sets of factors previously limited (arbitrarily) by an expert. Moreover, experts weighted factor relevancy on a continuous scale, which proved to be difficult in practice and may lead to unreliable results [24]. Most recently, Trendowicz et al. proposed an informal, integrated approach to selecting relevant productivity factors [24]. They used an analysis of existing project data in an interactive manner to support experts in identifying relevant factors for the purpose of effort estimation. Besides increased estimation performance, the factor selection contributed to increased understanding and improvement of software processes related to development productivity and cost.

## 3   An Integrated Factor Selection Method

In this paper, we propose an integrated method for selecting relevant productivity factors. The method employs a novel multi-criteria decision aid (MCDA) technique called *AvalOn* to combine the results of data- and expert-based factor selection.

### 3.1   Expert-based Factor Selection

Expert-based selection of relevant productivity factors is a two-stage process [24]. First, a set of candidate factors is proposed during a group meeting (brainstorming session). Next, factor relevancy criteria are identified and quantified on the Likert scale. Example criteria may include a factor's impact, difficulty, or controllability. *Impact* reflects the strength of a given factor's influence on productivity. *Difficulty* represents the cost of collecting factor-related project data. Finally, *controllability* represents the extent to which a software organization has an impact on the factor's value (e.g., a customer's characteristics are hardly controllable). Experts are then asked to individually evaluate the identified factors according to specified criteria.

### 3.2   Data-based Factor Selection

Data-based selection of relevant productivity factors employs one of the available factor weighting techniques. As compared to simple factor selection or ranking techniques, weighting provides experts with the relative distance between subsequent factors regarding their relevance. Selected weighting should be applicable to regression problems, i.e., to the continuous dependent variable (here: development productivity). Given the size of the factor space, optimal weighting [1] (small size) or weighting heuristics [11] (large size) should be considered. In this paper, we employ the *Regression ReliefF* (RRF) technique [16]. RRF is well suited for software engineering data due to its robustness against sparse and noisy data. The output of RRF (weighting) reflects the ratio of change to productivity explained by the input factors.

### 3.3 An Integrated Factor Selection Method

Integrated factor selection combines the results of data- and expert-based selections by means of the *AvalOn* MCDA method. It is the hierarchically (tree) structured model that was originally used in COTS (Commercial-of-the-shelf) software selection [15]. AvalOn incorporates the benefits of a tree-structured MCDA model such as the Analytic Hierarchy Process (AHP) [17] and leverages the drawbacks of pair-wise (subjective) comparisons. It comprises, at the same time, subjective and objective measurement as well as the incorporation of uncertainty under statistical and simulation aspects. In contrast to the AHP model, which only knows one node type, it distinguishes several node types representing different types of information and offering a variety of possibilities to process data. Furthermore, AvalOn offers a weight rebalancing algorithm mitigating typical hierarchy-based difficulties originating from the respective tree structure. Finally, it allows for any modification (add, delete) of the set of alternatives while maintaining consistency in the preference ranking of the alternatives.

#### 3.3.1 Mathematical background of the AvalOn method

As in many MCDA settings [25], a preference among the alternatives is processed by summing up *weight x preference* of an alternative. In AvalOn (Fig. 1), this is accomplished for the node types *root*, *directory*, and *criterion* by deploying the following abstract model in line with the meta-model structure:

$$pref_i(a) = \sum_{j \in subnodes(i)} w_j \cdot pref_j(a),$$

where *i* is a node in the hierarchy, *a* the alternative under analysis, *subnodes(i)* the set of child/sub-nodes of node *i*, $pref_j(a) \in [0..1]$ the preference of *a* in subnode *j*, and $w_j \in [0..1]$ the weight of subnode *j*. Hence $pref_i(a) \in [0..1]$.

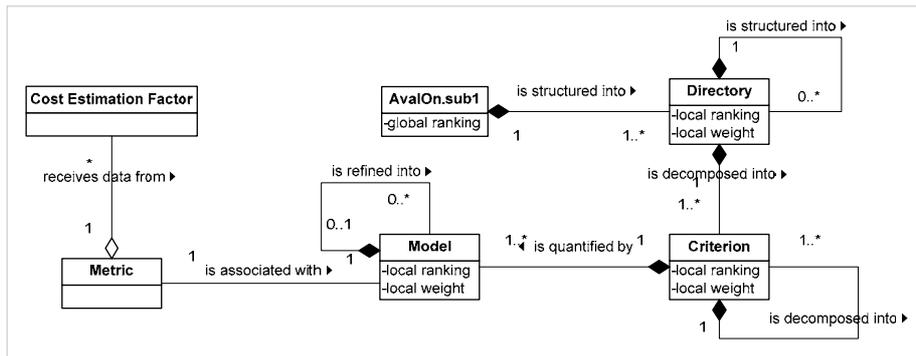

**Fig. 1.** Meta-model for factor selection

In each model, a value function (*val*) is defined, building the relation between data from {*metrics x alternatives*} and the assigned preference values. *val* may be defined almost in an arbitrary way, i.e., it allows for preference mappings of metric scaled data as well as categorical data.

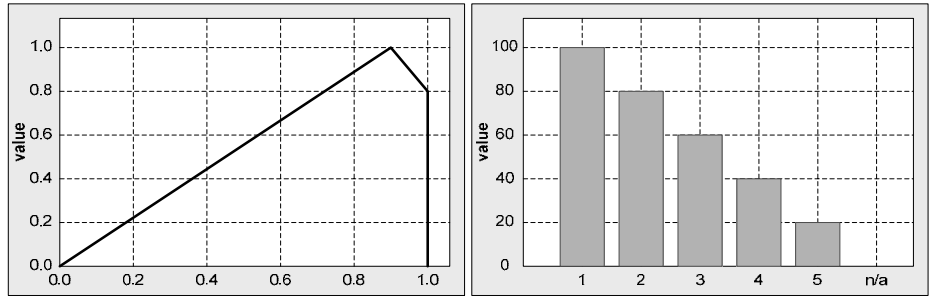

**Fig. 2.** Example *val* models

In this way, *val* can model the whole range of scales from semantic differential, via Likert to agreement scales. Please note that when calculating $pref_i(a)$ on the lowest criterion level, the direct outputs of the function *val* in the subnodes, which are *models* in this case, are weighted and aggregated. The full details of the general model definition for *val* is described in [18]. In this context, two examples for *val*, one metric scaled (figure on the left), and one categorical (figure on the right), are given in Fig. 2. On the x-axis, there are the input values of the respective metric, while the y-axis shows the individual preference output *val*. A full description of the AvalOn method can be found in [18, 15].

### 3.3.2 Application of the AvalOn method

AvalOn allows for structuring complex information into groups (element *directory* in the meta-model) and criteria (element *criterion* in the meta-model). Each directory as well as each criterion may be refined into sub-directories and sub-criteria. Each (sub)-criterion may then be refined into individual model(s) and sub-models. The models transform the measurement data coming from each alternative into initial preference values. The models providing the preferences based on each measurement by alternative are associated with a set of previously defined metrics. Bottom-up, the data coming from each alternative to be potentially selected (here: *productivity factors*) are then processed through the models and aggregated from there to the criteria and directory level(s). Finally, in the root node (here: *AvalOn.sub1*), the overall preference of the productivity factors based on their data about individual metrics is aggregated using a weighting scheme that is also spread hierarchically across the tree of decision (selection) criteria. The hybrid character of the setting in this paper can be modeled by combining expert opinion and objective data from, e.g., preliminary data analyses, into criteria and models within different directories, and defining an adequate weighting scheme.

## 4 Empirical Study

The integrated factor selection method proposed in this paper was evaluated in an industrial context. We applied the method for the purpose of software effort estimation and compared it with isolated expert- and data-based selection methods. Data-based factor selection employed the RRF technique [16] implemented in the WEKA data mining software [26]. Expert-based factor selection was performed as a multiple-expert ranking (see Section 4.2 regarding the ranking process).

### 4.1 Study Objectives and Hypotheses

The objective of the study was to evaluate in a comparative study expert- and data-based approaches and the integrated approach for selecting the most relevant productivity factors in the context of software effort estimation. For that purpose, we defined two research questions and related hypotheses:

*Q1. Do different selection methods provide different sets of productivity factors?*
- **H1.** Expert-based, data-based and integrated methods select different (probably partially overlapping) sets of factors.

*Q2. Which method (including not reducing factors at all) provides the better set of factors for the purpose of effort estimation?*
- **H2.** The integrated approach provides a set of factors that ensure higher performance of effort estimation than factors provided by expert- and data-based selection approaches when applied individually.

Some effort estimation methods such as stepwise regression [5] or OSR [4] already include embedded mechanisms to select relevant productivity factors. In our study, we wanted to evaluate in addition how preliminary factor selection done by an independent method influences the performance of such estimation methods. This leads us to a general research question:

*Q3. Does application of an independent factor selection method increase the prediction performance of an estimation method that already has an embedded factor selection mechanism?*

Answering such a generic question would require evaluating all possible estimation methods. This, however, is beyond the scope of this study. We limit our investigation to the OSR estimation method [4] and define a corresponding research hypothesis:
- **H3.** Application of an independent factor selection method does not increase the prediction performance of the OSR method.

Finally, in order to validate replicate the results of the most recent research regarding the application of data-based factor selection to analogy-based effort estimation (e.g., [6, 10]) we define the following question:

*Q4. Does application of a data-based factor selection method increase the prediction performance of an analogy estimation method?*
- **H4.** Application of a data-based factor selection method increases the prediction performance of a k-NN estimation method.

### 4.2 Study Context and Empirical Data

The empirical evaluation was performed in the context of Toshiba Information Systems (Japan) Corporation (TJSYS). The project measurement data repository contained a total of 76 projects from the information systems domain. Fig. 3 illustrates the variance of development productivity measured as function points (unadjusted, IFPUG) per man-month.

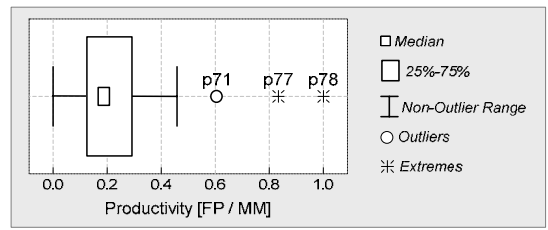

**Fig. 3.** Development productivity variance (data presented in a normalized form)

Expert assessments regarding the most relevant factors were obtained from three experts (see Table 1). During the group meeting (brainstorming session) an initial set of factors was identified. It was then grouped into project-, process, personnel-, and product-related factors as well as context factors. The first four groups refer to the characteristics of the respective entities (software project, development process, products, and stakeholders). The latter group covers factors commonly used to limit the context of software effort estimation or productivity modeling. The application domain, for instance, is often regarded as a context factor, i.e., an effort model is built for a specific application domain. Finally, experts were asked to select the 5 most important factors from each category and rank them from the most relevant (rank = 1) to least relevant (rank = 5).

**Table 1.** Experts participated in the study.

|  | **Expert 1** | **Expert 2** | **Expert 3** |
|---|---|---|---|
| Position/Role | Project manager | Developer | Quality manager |
| Experience [#working years] | 8 | 15 | 3 |
| Experience [#performed projects] | 30 | 15 | 40 |

### 4.3 Study Limitations

Unfortunately, the measurement repository available did not cover all relevant factors selected by the experts. It was also not possible to collect the data ex post facto. This prevented us from doing a full comparative evaluation of the three factor selection methods considered here for the purpose of software effort estimation. In order to at least get an indication of the methods' performance, we decided to compare them (instead of all identified factors) on the factors identified by experts for which measurement data were available. This would represent the situation where those factors cover all factors available in the repository and identified by experts.

### 4.4 Study Design and Execution

#### 4.4.1 Data Preprocessing

Measurement data available in the study suffered from incompleteness (44.3% missing data). An initial preprocessing was thus required in order to apply the data analysis techniques selected in the study. We wanted to avoid using simple approaches to handling missing data such as list-wise deletion or mean imputation, which significantly reduce data quantity and increase noise. Therefore, we decided to apply the k-Nearest Neighbor (k-NN) imputation method. It is a common *hot deck* method, in which k nearest projects minimizing a certain similarity measure (calculated on non-missing factors) are selected to impute missing data. It also proved to provide relatively good results when applied to sparse data in the context of software effort prediction [14]. Moreover, other more sophisticated (and potentially more effective) imputation methods required removing factor collinearities beforehand. Such a preprocessing step would, however, already be a kind of factor selection and might thus bias the results of the actual factor selection experiment. We adopted the k-NN imputation approach presented in [9]. In order to assure maximal performance of the imputation, before applying it, we removed factors and projects with large missing data ratio so that the total ratio of missing data was reduced to around one third, however, with minimal loss of non-missing data. We applied the following procedure: We first removed factors where 90% of the data were missing and next, projects where more than 55% of the data were still missing. As a result, we reduced the total rate of missing data to 28.8%, while losing a minimal quantity of information (removed 19 out of 82 factors and 3 out of 78 projects). The remaining 28.8% of missing data were imputed using the k-NN imputation technique.

#### 4.4.2 Empirical Evaluation

Let us first define the following abbreviations for the factor sets used in the study:

*FM*: factors covered by measurement data.
*$FM_R$*: relevant FM factors selected by the RReliefF method (factors with weight > 0)
*$FM_{R10}$*: the 10% most relevant $FM_R$ factors
*FE*: factors selected by experts
*FI*: factors selected by the integrated method
*FT*: all identified factors (FM∪FE)
*FC*: factors selected by experts for which measurement data are available (FM∩FE)
*$FC_{E25}$*: the 25% most relevant FC factors selected by experts
*$FC_{R25}$*: the 25% most relevant FC factors selected by the RRF method
*$FC_{I25}$*: the 25% most relevant FC factors selected by the integrated method

*Hypothesis H1.* In order to evaluate H1, we compared factor sets selected by the data-based, expert-based, and integrated method ($FM_R$, FE, and FI). For the 10 most relevant factors shared by all three factor sets, we compared the ranking agreement using Kendall's coefficient of concordance [20].

**Hypothesis H2.** In order to evaluate H2, we evaluated the estimation performance of two data-based estimation methods: k-Nearest Neighbor (k-NN) [19] and Optimized Set Reduction (OSR) [4]. We applied them in a leave-one-out cross validation on the following factor sets: FM, FC, $FC_{E25}$, $FC_{R25}$, and $FC_{I25}$.

**Hypothesis H3.** In order to evaluate H3, we compared the estimation performance of OSR (which includes an embedded, data-based factor selection mechanism) when applied on FM and $FM_{R10}$ factor sets.

**Hypothesis H4.** In order to evaluate H4, we compared the estimation performance of the k-NN method when applied on FM and $FM_{R10}$ factor sets.

To quantify the estimation performance in H2, H3, and H4, we applied the common accuracy and precision measures defined in [7]: magnitude of relative estimation error (MRE), mean and median of MRE (MMRE and MdMRE), as well as prediction at level 25% (Pred.25). We also performed an analysis of variance (ANOVA) [20] of MRE to see if the error for one approach was statistically different from another. We interpret the results as statistically significant if the results could be due to chance less than 2% of the time ($p < 0.02$).

## 5 Results of the Empirical Study

*Hypothesis H1:* Expert-based, data-based and integrated methods select different (probably partially overlapping) sets of factors.

After excluding the dependent variable (development productivity) and project ID, the measurement repository contained data on 61 factors. Experts identified a total of 34 relevant factors, with only 18 of them being already measured (FC). The RRF method selected 40 factors ($FM_R$), 14 of which were also selected by experts. The integrated approach selected 59 factors in total, with only 14 being shared with the former two selection methods. Among the FC factors, as many as 8 were ranked by each method within the top 10 factors (Table 2). Among the top 25% FC factors selected by each method, only one factor was in common, namely *customer commitment and participation*. There was no significant agreement (Kendall = 0.65 at p=0.185) between data- and expert-based rankings on the FC factors. The integrated method introduced significant agreement on ranks produced by all three methods (Kendall = 0.72 at p = 004).

*Interpretation (H1):* Data- and expert-based selection methods provided different (partially overlapping) sets of relevant factors. Subjective evaluation of the shared factors suggests that both methods vary regarding the assigned factor's importance; yet this could not be confirmed by statistically significant results. The integrated method introduced a consensus between individual selections (significant agreement) and as such might be considered as a way to combine the knowledge gathered in experts' heads and in measurement data repositories.

**Table 2.** Comparison of the ranks on FC factors (top 25% marked in bold)

| Productivity factor | $FC_E$ | $FC_R$ | $FC_I$ |
|---|---|---|---|
| Customer commitment and participation | **3** | **3** | **3** |
| System configuration (e.g., client-server) | 5 | **2** | 5 |
| Application domain (e.g., telecommunication) | **1** | 6 | **1** |
| Development type (e.g., enhancement) | 7 | **1** | 4 |
| Application type (e.g., embedded) | **2** | 7 | **2** |
| Level of reuse | 9 | **4** | 9 |
| Required product quality | 6 | 10 | 7 |
| Peak team size | 8 | 9 | 8 |

*Hypothesis H2: The integrated approach provides a set of factors that ensure higher performance of effort estimation than factors provided by expert- and data-based selection approaches when applied individually.*

A subjective analysis of the estimates in Table 3 suggests that the k-NN provided improved estimates when applied on a reduced FC factors set ($FC_{E25}$, $FC_{R25}$, and $FC_{I25}$), whereas OSR does not consistently benefit from independent factor reduction (by improved estimates). The analysis of the MRE variance, however, showed that the only significant ($p = 0.016$) improvement in estimation performance of the k-NN predictor was caused by the integrated factor selection method. The OSR predictor improved its estimates significantly ($p < 0.02$) only on the $FC_{E25}$ factors set.

*Interpretation (H2):* The results obtained indicate that a factors set reduced through an integrated selection contributes to improved effort estimates. Yet, this does not seem to depend on any specific way of integration. The k-NN predictor, which uses all input factors, improved on factors reduced by the AvalOn method. The OSR method, however, improved slightly on the factors reduced by experts. This interesting observation might be explained by the fact that OSR, which includes an embedded, data-based factor selection mechanism, combined this with prior expert-based factor selection. Still, the effectiveness of such an approach largely depends on the experts who determine (pre-select) input factors for OSR (expert-based selection is practically always granted higher priority).

**Table 3.** Comparison of various factor selection methods

| Predictor | *Factors Set* | MMRE | MdMRE | Pred.25 |
|---|---|---|---|---|
| k-NN | *FM* | 73.7% | 43.8% | 21.3% |
| | *FC* | 52.6% | 40.0% | 26.7% |
| | *$FC_{E25}$* | 46.3% | 38.5% | 33.3% |
| | *$FC_{R25}$* | 48.3% | 36.9% | 29.3% |
| | ***$FC_{I25}$*** | **47.5%** | **33.3%** | **30.7%** |
| OSR | *FM* | 59.7% | 50.8% | 17.3% |
| | *FC* | 65.9% | 59.2% | 18.7% |
| | ***$FC_{E25}$*** | **30.7%** | **57.9%** | **24.0%** |
| | *$FC_{R25}$* | 66.2% | 52.1% | 14.7% |
| | *$FC_{I25}$* | 65.1% | 57.9% | 14.7% |

***Hypothesis H3:*** *Application of an independent factor selection method does not increase the prediction performance of the OSR method.*

A subjective analysis of OSR's estimation error (Table 3 and Table 4) suggests that it performs generally worse when applied on the factors chosen by an independent selection method. This observation was, however, not supported by the analysis of the MRE variance. The exception was the FC set reduced by experts ($FC_{E25}$), on which a slight, statistically significant improvement of the OSR's predictions was observed.

**Table 4.** Results of data-based factor selection

| Predictor | Factors Set | MMRE | MdMRE | Pred.25 | ANOVA |
|---|---|---|---|---|---|
| k-NN | FM | 73.7% | 43.8% | 21.3% | p = 0.39 |
| | $FM_{R10}$ | 56.8% | 40.7% | 22.7% | |
| OSR | FM | 59.7% | 50.8% | 17.3% | p = 0.90 |
| | $FM_{R10}$ | 68.1% | 59.1% | 16.0% | |

***Interpretation (H3).*** The results obtained indicate that no general conclusion regarding the impact of independent factor selection on the prediction performance of OSR can be drawn. Since no significant deterioration of estimation performance was observed, application of OSR on the reduced set of factors can be considered useful due to the reduced cost of measurement. Yet, improving OSR's estimates might require a selection method that is more effective than the selection mechanism embedded in OSR.

***Hypothesis H4:*** *Application of a data-based factor selection method increases the prediction performance of a k-NN estimation method.*

A subjective impression of improved estimates provided by the k-NN predictor (Table 4) when applied on the reduced factors set ($FM_{R10}$) was, however, not significant in the sense of different variances of MRE (p = 0.39). Yet, estimates provided by the k-NN predictor improved significantly when used on the FC data set reduced by the integrated selection method (p = 0.016). The two individual selection methods did not significantly improve performance of the k-NN predictor.

***Interpretation (H4):*** Although a subjective analysis of the results (Table 3 and Table 4) suggests improved estimates provided by the k-NN predictor when applied on reduced factors sets, no unambiguous conclusion can be drawn. The performance of k-NN improved significantly only when applied on factors identified from the FC set by the integrated selection method (the $FC_{I25}$ set). This might indicate that k-NN's performance improvement depends on the applied factor selection method (here, the integrated method was the best one).

**Threats to Validity**

We have identified two major threats to validity that may limit the generalizability of the study results. First, the estimation performance results of the factor selection methods investigated, compared on the FC set, may not reflect their true characteristics, i.e., as compared on the complete set of identified factors (threat to hypothesis H2). Yet, a lack of measurement data prevented us from checking on this. Second, the RRF method includes the k-NN strategy to search through the factor space and iteratively modify factor weights. This might bias the results of k-NN-based estimation by contributing to better performance of k-NN (as compared to OSR) on factors selected by RRF (threat to hypotheses H3 and H4).

## 6  Summary and Further Work

In this paper, we proposed an integrated approach for selecting relevant factors influencing software development productivity. We compared the approach in an empirical study against selected expert- and data-based factor selection approaches.

The investigation performed showed that expert- and data-based selection methods identified different (only partially overlapping) sets of relevant factors. The study indicated that the *AvalOn* method finds a consensus between factors identified by individual selection methods. It combines not only the sets of relevant factors, but the individual relevancy levels of selected factors. We showed that in contrast to data- and expert-based factor selection methods, the integrated approach may significantly improve the estimation performance of estimation methods that do not include an embedded factor selection mechanism. Estimation methods that include such a mechanism may, however, benefit from integrating their capabilities with expert-based factor selection.

The study did not replicate the observation of similar investigations regarding improved estimation performance on the factor sets reduced by a data-based selection method. Neither of the estimation methods employed in the study (k-NN and OSR) improved significantly when applied on factor sets reduced by the RReliefF method. Although k-NN improved in terms of aggregated error measures (e.g., MMRE) the difference in the MRE variance was insignificant. The results obtained for the OSR method may indicate that the change of its prediction performance when applied on a reduced set of factors depends on the selection method used.

Finally, we also observed that the function point adjustment factor (FPAF) was not considered among the most relevant factors, although factor selection was driven by a variance on development productivity calculated from unadjusted function point size. Moreover, some of the factors considered as relevant (e.g., performance requirements) belong to components of the FPAF. This might indicate that less relevant sub-factors of the FPAF and/or the adjustment procedure itself may hide the impact of relevant factors. Considering sub-factors of FPAF individually might therefore be more beneficial.

In conclusion, factor selection shall be considered as an important aspect of software development management. Since individual selection strategies seem to provide inconsistent results, integrated approaches should be investigated to support software practitioners in limiting the cost of management (data collection and analysis) and increasing the benefits (understanding and improvement of development processes).

Further work will focus on a full evaluation of the three selection strategies presented on a complete set of measurement data (including data on all factors identified by experts). Finally, methods to identify and explicitly consider factor dependencies require investigation. Such information may not only improve the performance of effort estimation methods, but also the understanding of interactions among organizational processes influencing development productivity.

**Acknowledgments.** We would like to thank Toshiba Information Systems (Japan) Corporation and all involved experts who greatly contributed to the study. We would also like to thank the Japanese Information-technology Promotion Agency for supporting the study. Finally, we would like to thank Sonnhild Namingha and Marcus Ciolkowski for reviewing the paper.

# References

1. Auer, M., Trendowicz, A., Graser, B. Haunschmid, E. Biffl, S.: Optimal project feature weights in analogy-based cost estimation: improvement and limitations. IEEE Transactions on Software Engineering, Vol. 32, No. 2. (2006) 83–92.
2. Boehm, B.W., Abts, C., Brown, A.W., Chulani, S., Clark, B.K., Horowitz, E., Madachy, R., Refer, D., Steece B.: Software Cost Estimation with COCOMO II. Prentice Hall (2000).
3. Boehm, B.W.: Software Engineering Economics. Prentice-Hall (1981).
4. Briand, L., Basili, V., Thomas, W.: A Pattern Recognition Approach for Software Engineering Data Analysis. IEEE Transactions on Software Engineering, Vol. 18, No. 11. (1992) 931–942.
5. Chatterjee, S., Hadi, A.S., Price, B.: Regression Analysis by Example (3rd Edition). Wiley-Interscience. (1999).
6. Chen, Z., Menzies, T., Port, D., Boehm, B.: Finding the right data for software cost modeling. IEEE Software, Vol. 22, No. 6. (2005) 38–46.
7. Conte S., Dunsmore H., Shen V.Y.: Software Engineering Metrics and Models. CA: Benjamin Cummings. (1986).
8. Guyon, I., Elisseeff, A.: An Introduction to Variable and Feature Selection. Journal of Machine Learning Research, Vol. 3. (2003) 1157–1182.
9. Jönsson, P., Wohlin, C.: An Evaluation of k-Nearest Neighbour Imputation Using Likert Data. Proc. 10th Int'l Symposium on Software Metrics. (2005) 108–118.
10. Kirsopp, C., Shepperd, M., Hart, J.: Search Heuristics, Case-based Reasoning and Software Project Effort Prediction. Proc. Genetic and Evolutionary Computation Conference. (2002) 1367–1374.
11. Li, J., Ruhe, G.: A comparative study of attribute weighting heuristics for effort estimation by analogy. Proc. Int'l Symposium on Empirical Software Engineering. (2006) 66–74.
12. Liang, T., Noore, A.: Multistage software estimation." Proc. 35th Southeastern Symposium on System Theory. (2003) 232–236.
13. Maxwell, K.D., Van Wassenhove, L., Dutta, S.: Software development productivity of European space, military, and industrial applications. IEEE Transactions on Software Engineering, Vol. 22, No. 10. (1996) 706–718.
14. Myrtveit, I., Stensrud, E., Olsson, U.H.: Analyzing data sets with missing data: An empirical evaluation of imputation methods and likelihood-based methods. IEEE Transactions Software Engineering, Vol. 27. (2001) 999–1013.
15. Ochs, M., Pfahl, D., Chrobok-Diening, G., Nothhelfer-Kolb B.: A Method for efficient measurement-based COTS Assessment & Selection – method description and evaluation results. Proc. 7th Int'l Software Metrics Symposium. (2001).
16. Robnik-Sikonja, M., Kononenko, I.: Theoretical and Empirical Analysis of ReliefF and RReliefF. The Machine Learning Journal, Vol. 53. (2003) 23–69.
17. Saaty, T.L.: The Analytic Hierarchy Process. McGraw-Hill, New York (1990).

18. Schillinger, D.: Entwicklung eines simulationsfähigen COTS Assessment und Selection Tools auf Basis eines für Software adequaten hierarchischen MCDM Meta Modells. Ms Thesis, Dept. of Computer Science, TU Kaiserslautern, (Supervisors): Prof. Dr. D. Rombach, Michael Ochs, Kaiserslautern, Germany (2006).
19. Shepperd, M., Schofield, C.: Estimating Software Project Effort Using Analogies. IEEE Transactions on Software Engineering, Vol. 23, No. 12. (1997) 736–743.
20. Sheskin, D.J. Sheskin, D.: Handbook of Parametric and Nonparametric Statistical Procedures (2$^{nd}$ Edition). Chapman & Hall/CRC (2000).
21. Spector, P.: Summated Rating Scale Construction. Sage Publications (1992).
22. Subramanian, G.H., Breslawski, S.: Dimensionality reduction in software development effort estimation. Journal of Systems and Software, Vol. 21, No. 2. (1993) 187–196.
23. The Standish Group. CHAOS Chronicles. West Yarmouth, MA (2003).
24. Trendowicz, A., Heidrich, J., Münch, J., Ishigai, Y., Yokoyama, K., Kikuchi, N.: Development of a Hybrid Cost Estimation Model in an Iterative Manner. Proc. 28$^{th}$ Int'l Conference on Software Engineering, Shanghai, China (2006) 331–340.
25. Vincke, P. "Multicriteria Decision-aid", John Wiley & Sons: Chichester, 1992.
26. Witten, I.H., Frank, E.: Data Mining: Practical machine learning tools and techniques (2$^{nd}$ Edition). Morgan Kaufmann, San Francisco (2005).
27. Trendowicz, A.: Factors Influencing Software Development Productivity - State of the Art and Industrial Experiences. Report no. 008.07/E. Fraunhofer IESE, Kaiserslautern, Germany (2007).